\documentclass{article}

\usepackage{PRIMEarxiv}
\usepackage{amsmath}
\usepackage{cite}
\usepackage{amsmath,amssymb,amsfonts}
\usepackage{algorithmic}
\usepackage{graphicx}
\usepackage{siunitx}
\usepackage{textcomp}
\usepackage{upgreek}
\usepackage[utf8]{inputenc} 
\usepackage[T1]{fontenc}    
\usepackage{hyperref}       
\usepackage{url}            
\usepackage{booktabs}       
\usepackage{amsfonts}       
\usepackage{nicefrac}       
\usepackage{microtype}      
\usepackage{lipsum}
\usepackage{fancyhdr}       
\usepackage{graphicx}       
\graphicspath{{media/}}     

\title{Directional dark field retrieval with single-grid x-ray imaging}

\author{
 Michelle K Croughan \textsuperscript{1}, Ying Ying How \textsuperscript{1}, Allan Pennings\textsuperscript{2}, Kaye S. Morgan\textsuperscript{1} \\
 \\
\textsuperscript{1}
School of Physics and
Astronomy, Monash University, Clayton, VIC 3800, Australia\\
\textsuperscript{2}
Optical Sciences Centre, Swinburne University of Technology, Melbourne, VIC 3122, Australia
\\
\texttt{email: Michelle.Croughan@monash.edu}}

\begin{document}
\maketitle
© 2023 Optical Society of America. One print or electronic copy may be made for personal use only. Systematic reproduction and distribution, duplication of any material in this paper for a fee or for commercial purposes, or modifications of the content of this paper are prohibited.
\newline

\begin{abstract}
Directional dark-field imaging is an emerging x-ray modality that is sensitive to unresolved anisotropic scattering from sub-pixel sample microstructures. A single-grid imaging setup can be used to capture dark-field images by looking at changes in a grid pattern projected upon the sample. By creating analytical models for the experiment, we have developed a single-grid directional dark-field retrieval algorithm that can extract dark-field parameters such as the dominant scattering direction, and the semi-major and -minor scattering angles. We show that this method is effective even in the presence of high image noise, allowing for low-dose and time-sequence imaging.
\end{abstract}


\section{Introduction}

The field of x-ray imaging research aims to develop new methods that either extract new information about the sample, or extract sample information with less radiation exposure than existing methods. A recent example is the development of various techniques that use x-ray phase contrast to detect weakly-attenuating sample features~\cite{endrizzi2018x}. However, our ability to observe sample structures at small spatial scales is still typically limited by the spatial resolution of the imaging system. A potential solution is the x-ray dark-field signal, generated by unresolved scattering from sample microstructures. Directly resolving these structures would require smaller effective pixel sizes and increased flux resulting in higher radiation exposure to the sample. Measuring the dark field allows us access to sub-pixel information, without applying the radiation dose that would generally be required for direct imaging of those sample features. 

\begin{figure}[htbp]
\centering\includegraphics[width = 4.9in]{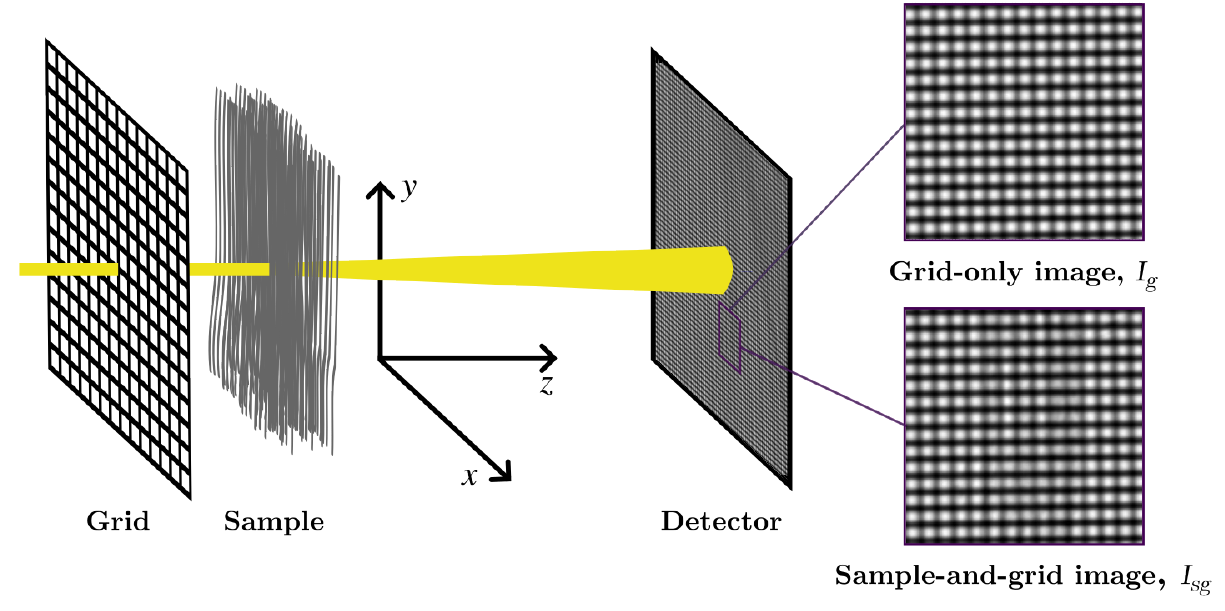}
\caption{An illustration of a single-grid imaging experiment setup. An x-ray beamlet is shown propagating along the $z$ axis from left to right. This beamlet first travels through the reference pattern, in this case a grid, then through a sample which introduces a dark-field signal via scattering from the unresolved sample microstructures. The x-ray wavefield intensity is then recorded on the detector some distance downstream of the sample. Inserts of grid-only and sample-and-grid images are shown on the right. Note the subtle variation in grid visibility in the sample-and-grid image, primarily observed as a subtle blurring along the horizontal axis of the image reducing the visibility of the vertical grid lines. This is due to a combination of attenuation (reducing the local intensity), resolved phase affects (shifting the grid pattern) and the dark field (locally blurring the grid pattern). The experimental coordinate system, $(x,y,z)$, is shown.}
\label{fig: experimental set up}
\end{figure}

Refractive properties of the sample introduce phase shifts that result in the displacement of the x-ray intensity at the image plane. When the coherent x-ray wavefield is imaged some distance downstream from the sample this produces interference fringes around material interfaces. Sub-pixel sample structures produce small phase shifts that cannot be directly resolved, but many of these structures together will alter the wavefield coherence. Very fine microstructures will also cause divergence of the x-ray beam by inducing small-angle or ultra-small x-ray scattering (SAXS/USAXS). This can be collectively modelled as a scattering of the x-ray beam that smooths out local intensity variations in an image. This is shown in Fig.~\ref{fig: experimental set up} where the incoming x-ray beamlet diverges after passing through the sample. This blurring of the observed intensity due to the presence of sub-pixel structures is what we define as the x-ray dark field. Various experimental techniques have been previously used to reconstruct a dark-field signal, the nature of which can vary between set-ups. In the case of a grating interferometrey setup, changes in the visibility of a stepping curve is used as a measure of the dark-field strength~\cite{pfeiffer_hard-x-ray_2008,Jensen_2010_A}. Similarly single-grid/speckle-tracking type techniques have used the reduction in reference pattern visibility~\cite{zdora_x-ray_2017,How:22}. Analyser-based imaging can be used to measure the dark-field angle over which a local region of the sample scatters the beam~\cite{pagot2003, rigon2007_abi_3}, as can edge illumination~\cite{endrizzi2014}. When combining grating interferometry with computed tomography the dark-field effect is quantified by a linear diffusion coefficient~\cite{bech2010quantitative}. 

While most x-ray dark-field imaging techniques assume isotropic scattering it is possible to have directional scattering. This is seen when the sample scatters the x-ray beam more in some directions relative to other directions. This anisotropic scattering is known as `directional dark field'~\cite{Jensen_2010_A, jensenDirectionalXrayDarkfield2010_B}. Aligned elongated microstructures in a sample will create a dark-field signal that is stronger perpendicular to the direction of the structures, and weaker in the direction parallel to those structures. This anisotropic scattering can be observed as an asymmetrical dark-field signal, with varying strength over different angles within the image plane. Extracting directional dark-field information about a sample was first shown using grating interferometry by taking multiple data sets with the sample oriented in different positions relative to the grating lines~\cite{Jensen_2010_A,jensenDirectionalXrayDarkfield2010_B}. This approach can also be combined with computed tomography to retrieve 3D directional dark-field volumes~\cite{jud_dentinal_2016}. Alternatively, Moire fringes produced in a stationary grating interferometry setup with trochoidal motion of a sample can be analysed to determine the orientation of microstructures~\cite{sharmaTrochoidalXrayVector2018}. A single phase grating with an array of circular features has also been used to collect directional information with one sample exposure~\cite{kagias2DOmnidirectionalHardXRayScattering2016} or with multiple exposures to achieve directional dark-field computed tomography~\cite{kimXrayScatteringTensor2020}. Extracting the directional dark-field information allows us to determine the orientation of microstructures in a sample without directly resolving them. 

Single-grid x-ray imaging is a phase contrast technique that can extract both attenuation and differential-phase signals~\cite{wen2010single, morgan2011quantitative}. An absorption grid~\cite{bennettGratingbasedSingleshotXray2010, morgan2011quantitative,wen2010single}, phase grid~\cite{morganSensitiveXrayPhase2013, rizzi2013,groenendijk2020, riedel2021, gustschin2021}, or irregular speckle reference pattern~\cite{morganXrayPhaseImaging2012,berujonTwoDimensionalXRayBeam2012,zanette2014, zdoraStateArtXray2018, smith2022} is placed a given distance upstream of the detector so that the intensity variations from the pattern are directly resolved. In some cases, multiple sample exposures are taken with the grid/speckle in multiple positions to increase the spatial resolution of the final phase~\cite{berujon2012, zdora_x-ray_2017, riedel2021}, dark-field~\cite{berujon2012, berujon2015, zdora_x-ray_2017}, or more-recently, directional dark-field image\cite{smith2022}. While multiple sample exposures increase the resolution of the data extracted, they require the sample to remain static for an extended period of time. In the case where this is not possible, a single-exposure technique would allow for time-resolved imaging of moving samples.

The dark-field signal retrieved from the above techniques is qualitative in that it is generally expressed as a reduction in visibility specific to a given experimental set up, and is not converted into a quantitative measure of the sample's scattering that can be used to compare different samples in different experimental conditions. In this paper we focus on a single-exposure technique that retrieves quantitative measures of the sample's scattering angle from the dark-field signal. First a grid-only image, $I_g$, is collected, then the sample is placed into the x-ray beam, close to the grid, to obtain the sample-and-grid image, $I_{sg}$. A diagram of this experimental setup is shown in Fig.~\ref{fig: experimental set up}. Sample attenuation reduces the local intensity of the reference pattern, while transverse shifts of the reference pattern can provide measurements of phase shifts by comparing $I_{g}$ to $I_{sg}$.  

Several methods exist to analyse $I_g$ and $I_{sg}$ to also extract a qualitative dark-field image from this blurring~\cite{wen2010single,morganSensitiveXrayPhase2013, zhou2018, dreier2020, How:22}. Recently, it was shown by How et. al.~\cite{How:22} that a quantitative dark-field strength can be extracted from single-grid or speckle-based images by selecting a small window in each of $I_g$ and $I_{sg}$ and comparing via cross-correlation. To do this, a model is fitted to the cross-correlation images to determine the angle over which the microstructures have isotropically scattered the x-ray beam~\cite{How:22}. Here we will build on this method to also extract scattering asymmetry and directional dark-field information. As single-grid imaging only requires one sample exposure, it can be used for low-dose and fast dynamic imaging which can be advantageous for dose sensitive and moving samples such as biological tissues or items moving along a conveyor belt. 

Current methods for dark-field imaging may extract an anisotropic signal, use a single exposure or retrieve a  quantitative measure, however no existing method does all three. The aim for this paper is to combine these strengths into one technique in order to obtain a quantitative measure of the directional dark field in a single exposure. This is achieved by creating analytical models that describe the experimental images. We use these models to predict how the dark-field signal appears in the sample-and-grid images. With this basis, we developed a novel single-grid directional dark-field retrieval algorithm that allows us to extract the dark-field parameters from a single sample exposure.

\section{Analytical model}
The localised blurring of a grid due to the x-ray dark field generated by a sample can be described by
\begin{equation}
    I_{sg}(x,y) = I_{g}(x,y) \otimes DF(x,y),
	\label{eq: dark-field model}
\end{equation}
where $I_{sg}$ is the convolution, $\otimes$, of $I_g$ with a position-dependent dark-field scattering kernel, $DF$, that describes the local blurring due to the dark-field signal~\cite{jensenDirectionalXrayDarkfield2010_B}. Within this section of the manuscript, we present a widely-applicable model based on a local-sinusoidal intensity variation. The subsequent algorithm could be rederived using a different expression for $I_g$ to describe a particular experiment, for example if using circular gratings~\cite{ kagias2DOmnidirectionalHardXRayScattering2016,kimXrayScatteringTensor2020}. However, since the local cross-correlation we use to compare $I_{sg}$ to $I_{g}$ (Sec.~\ref{sec:CrossCorr}) does not significantly change shape for different input patterns, we believe the algorithm may be applied to grids not strictly described by the model presented in Sec.~\ref{sec: grid-only-image}~\cite{How:22}. The following section describes the model used for the dark-field scattering kernel which is convolved with the grid-only image model (Sec.~\ref{sec: grid-only-image}) as per Eq.~\ref{eq: dark-field model} to define the sample-and-grid image model (Sec.~\ref{sec: sample-and-grid image}). 

\subsection{Dark-field model}
\begin{figure}[htbp]
\centering\includegraphics[width = 4.9in]{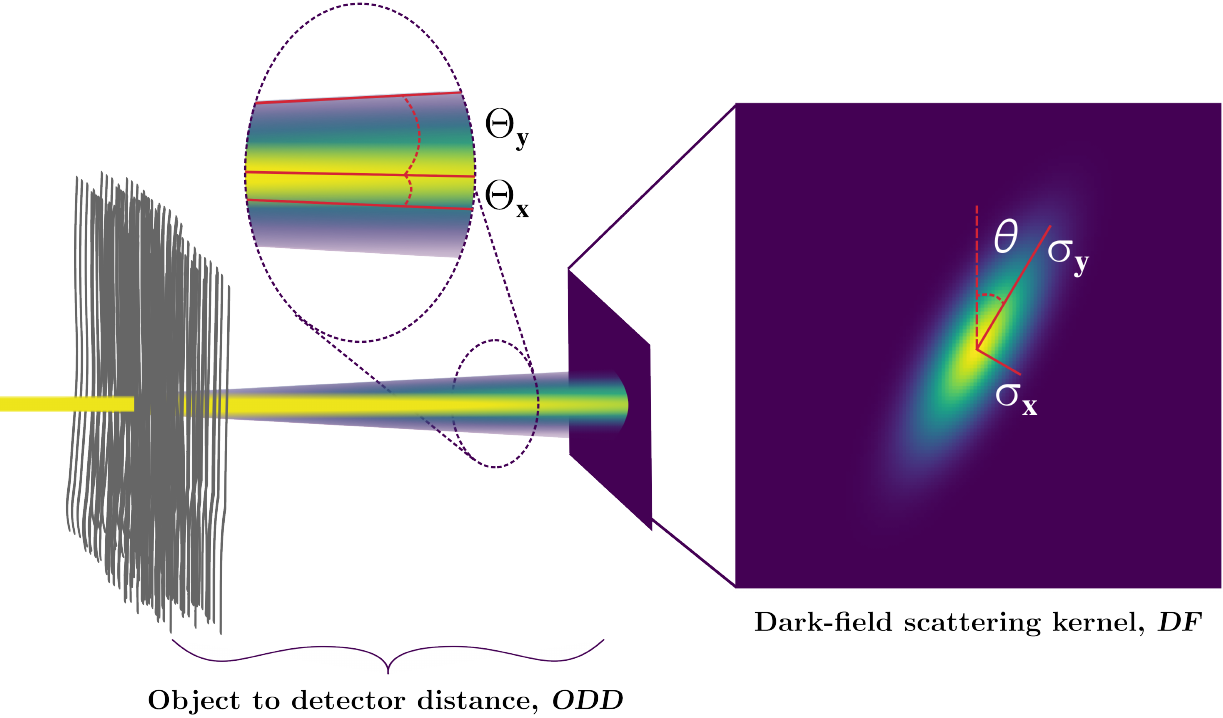}
\caption{A schematic demonstrating the scatter of an x-ray beamlet by the microstructures within a sample as the beamlet propagates from left to right. A simulated example of a dark-field scattering kernel, $DF$, is shown on the right with parameters $\theta = \pi/6$ radians, $\sigma_x = 1$ and $\sigma_y=3$ at a distance $ODD$ from the object (units of length are arbitrary here). The size of this kernel is dependent on the position of the detector since the beamlet expands as it propagates past the sample. The scattering cone half-angles, $\Theta_x$ and $\Theta_y$, quantify the divergence of the x-ray beamlet in a way that is independent of the object-to-detector distance.}
\label{fig: dark field model}
\end{figure}

The dark-field scattering kernel, $DF$, describes the local blurring at a given distance resulting from the divergence of an x-ray beamlet at the image plane as shown in Fig.~\ref{fig: dark field model}. It can be modelled as a two-dimensional normalised asymmetrical Gaussian centred at the origin using the following expressions
\begin{equation}
   DF(x,y) =  \frac{1}{2 \pi \sigma_x \sigma_y} \exp \left[ -Ax^2 - 2Bxy - Cy^2 \right],
   \label{eq: dark field kernel}
\end{equation}
\begin{equation}
    A = \frac{\cos ^2 \theta}{2 \sigma_x^2} + \frac{\sin ^2 \theta}{2 \sigma_y^2},
\end{equation}
\begin{equation}
    B = - \frac{\sin 2 \theta}{4 \sigma^2_x} + \frac{\sin 2 \theta}{4\sigma^2_y},
\end{equation}
\begin{equation}
    C = \frac{\sin^2 \theta}{2 \sigma^2_x} + \frac{\cos^2 \theta}{2\sigma^2_y},
\end{equation}
where $\sigma_x$ and $\sigma_y$ are the width of the blurring kernel (quantified by the Gaussian standard deviation) in the $x$ and $y$ directions, and $\theta$ describes the rotation within the $xy$ image plane in a clockwise direction~\cite{jensenDirectionalXrayDarkfield2010_B}. As shown in Fig.~\ref{fig: dark field model}, the dark-field scattering kernel is dependent on the object-to-detector distance, $ODD$. At greater distances $\sigma_x$ and $\sigma_y$ will become larger, so to obtain an $ODD$ independent measure they can be converted into scattering cone half-angles. Using the small angle approximation, we define the scattering angles as
\begin{equation}
    \Theta_x = \frac{\sigma_x}{ODD},
\end{equation}
\begin{equation}
    \Theta_y = \frac{\sigma_y}{ODD}.
\end{equation}
Both $\Theta_x$ and $\Theta_y$ are set-up independent measures of the dark-field strength that can be determined from the dark-field scattering kernel. As they are derived from the standard deviation they describe the cone half-angle in which $\sim 68.2\%$ of the initial x-ray beamlet intensity is contained, making this a quantitative measure. Note that the preceding isotropic work from How et al. used the full cone angle $2\Theta$~\cite{How:22}.

Depending on the analysis required one might want to use a more complex function to model the dark field~\cite{modregger2012, berujon2015}, but for most situations a Gaussian model will be suitable and results in a robust solution. Extracting the parameters $\theta$, $\sigma_x$ and $\sigma_y$ from experimental data will allow us to fully describe the dark-field signal.

\subsection{Grid-only image model} \label{sec: grid-only-image}

This model describes the intensity observed at the image plane with only the grid present in the x-ray beam. The grid produces an image with equidistant horizontal and vertical grid lines of constant thickness. Typically for single-grid imaging the grid period, $p$, is 8 to 16 pixels in order to produce a high visibility intensity pattern, but not so widely-spaced that fine sample features are lost~\cite{morganSensitiveXrayPhase2013, macindoe2016requirements, zdoraStateArtXray2018}. As the grid will be blurred by source size blurring and the point spread function of the imaging system, we can use sine waves to describe the intensity pattern, even when using a relatively `square' grid. The period of the sinusoidal model will be $p$, and in the case of an attenuation grid, the amplitude will relate to the x-ray absorption of the grid material, $\alpha$, where $\alpha$ can be between $0$ and $1$ (no absorption or complete absorption of the beam respectively). In the case of a phase grid, $\alpha$ would describe the difference between the minimum and maximum values of the intensity image and the factor of $3/4 \alpha$ in Eq.~\ref{eq:constant grid lines} would be removed.  The 2D intensity pattern seen in the grid-only image can be described by
\begin{equation}
    I_g(x,y) = 1 - \frac{3}{4}\alpha + \frac{1}{4}\alpha \sin\left(\frac{2\pi}{p}x\right) + \frac{1}{4}\alpha \sin\left(\frac{2\pi}{p}y\right) + \frac{1}{8} \alpha \cos \left(\frac{2\pi}{p} (x-y) \right) -\frac{1}{8} \alpha \cos \left(\frac{2\pi}{p} (x+y) \right).
    \label{eq:constant grid lines}
\end{equation}
A simulated example of this expression is shown in Fig.~\ref{fig: grid models}(a), plotted as a function of $x$ and $y$ across the image plane.

\subsection{Sample-and-grid image model}\label{sec: sample-and-grid image}

The sample can alter the intensity pattern via the dark field, sample transmission and by introducing resolvable phase changes that alter the transverse position of the intensity pattern at the image plane. The sample transmission, $T$, scales the intensity where $T$ can be between $0$ and $1$ (no transmission or complete transmission of the beam through the sample respectively). We make the assumption that resolved shifts in the intensity pattern due to phase effects are typically less than one grid period which is true in a well-chosen experimental setup~\cite{macindoe2016requirements}. Thus, as the intensity pattern shifts are small enough to not significantly displace the grid pattern we do not attempt to correct these shifts in the raw data but rather include them in the correlation image model as discussed in Sec.~\ref{sec:CrossCorr}. Larger, localised phase affects such as propagation-based phase fringes that are similar to the grid period in size can alter the attenuation signal. However, they are not accounted for in this model potentially producing artefacts in the retrieved images, see Sec.~\ref{sec: phase results} for further discussion. 

\begin{figure}[htbp]
\centering\includegraphics[width = 4.9in]{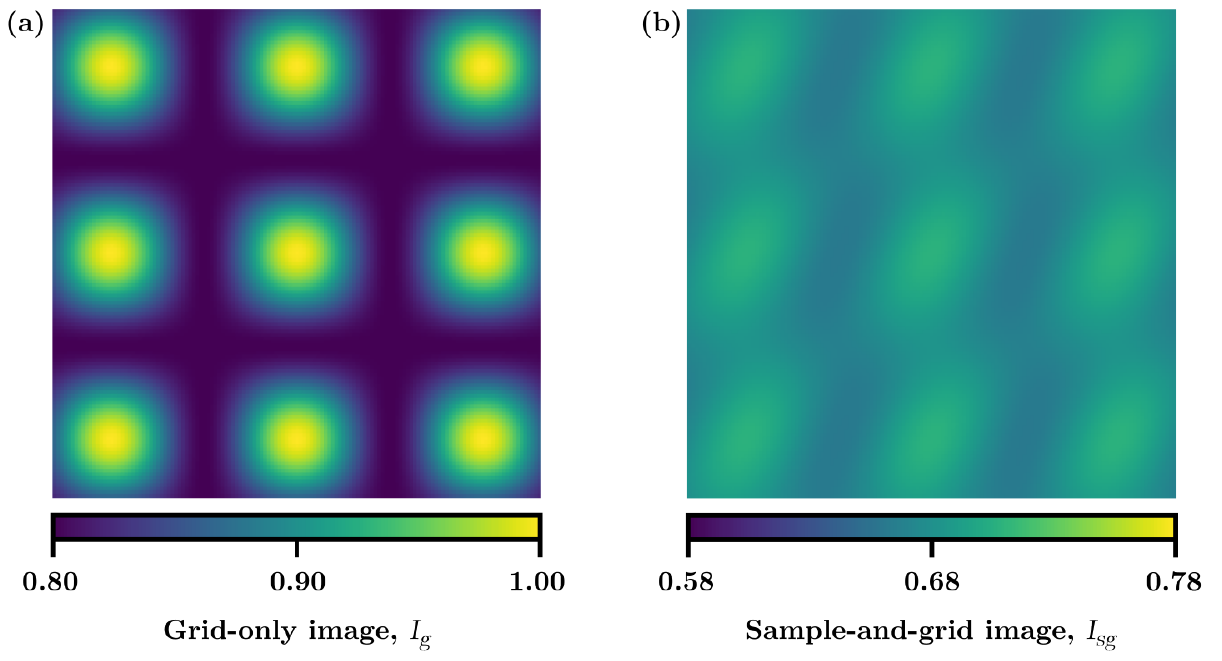}
\caption{Simulated examples of the analytical models for $I_g$ and $I_{sg}$. (a) Grid-only image, $I_g$, with parameters grid period $p=8$, and material absorption $\alpha = 0.2$. (b) Sample-and-grid image, $I_{sg}$, which is the convolution of $I_g$ (a) and $DF$ (Fig.~\ref{fig: dark field model})  scaled by the sample transmission of $T = 0.8$. Units of length are arbitrary. Note that $I_{sg}$ has lower visibility and has been blurred by an asymmetrical dark-field scattering kernel.}
\label{fig: grid models}
\end{figure}

Substituting the model for the grid image (Eq.~\ref{eq:constant grid lines}, Fig.~\ref{fig: grid models}(a)) and dark-field scattering kernel (Eq.~\ref{eq: dark field kernel}, Fig.~\ref{fig: dark field model}) into Eq.~\ref{eq: dark-field model} and scaling by the sample transmission factor of $T$ we are able to describe the sample-and-grid image model as
\begin{equation} \label{Eq: analytical expression for Isg}
\begin{split}
      I_{sg}(x,y) =T \Bigg( & 1 - \frac{3}{4}\alpha + \frac{1}{4}\alpha e^{-a - b\cos\left(2\theta\right)} \sin\left(\frac{2 \pi}{p}x\right) + \frac{1}{4}\alpha e^{-a+b\cos\left(2\theta\right)} \sin\left(\frac{2 \pi}{p}y\right)  \\  & +\frac{1}{8} \alpha e^{-2a-2b\sin\left(2\theta\right)} \cos\left(\frac{2\pi}{p}(x-y)\right) - \frac{1}{8} \alpha e^{-2a+2b\sin\left(2\theta\right)} \cos\left(\frac{2\pi}{p}(x+y)\right) \Bigg),
\end{split}
\end{equation}
where
\begin{equation}
    a = \frac{\pi^2}{p^2} \left(\sigma_x^2 + \sigma_y^2\right),
    \label{eq: a}
\end{equation}
\begin{equation}
    b = \frac{\pi^2}{p^2} \left(\sigma_x^2 - \sigma_y^2\right).
    \label{eq: b}
\end{equation}
The variables; $T$, $\theta$, $\sigma_x$ and $\sigma_y$ are all dependent on the location within the image plane. Fig.~\ref{fig: grid models}(b) gives an example of what the resulting sample-and-grid image looks like if the dark-field scattering kernel shown in Fig.~\ref{fig: dark field model} was constant across the whole image plane.

\subsection{Auto- and cross-correlations}
\label{sec:CrossCorr}
In order to extract the parameters of the dark-field scattering kernel we need to assess how the sample-and-grid image has changed relative to the grid-only image. It is possible to directly compare the amplitudes of the matching trigonometric terms in Eq.~\ref{eq:constant grid lines} and~\ref{Eq: analytical expression for Isg}, however cross-correlation analysis is an established technique that provides noise-robust results~\cite{morganSensitiveXrayPhase2013,How:22}. Single exposure grid/speckle-tracking phase retrieval has been achieved using cross-correlation analysis to detect if the reference pattern has been locally shifted~\cite{morgan2011quantitative,berujonTwoDimensionalXRayBeam2012}. Cross-correlation analysis has also been used in single-grid dark-field imaging to measure the isotropic scattering angle~\cite{How:22}. Making the comparison in correlation space is beneficial as it is less susceptible to error introduced by image noise and can be used for non-regular grids. To solve for both the transmission, $T$, and the dark-field scattering kernel, $DF$, we compute the cross-correlation of $I_g$ and $I_{sg}$, and also the auto-correlation of $I_g$. Phase shifts can displace the grid pattern resulting in a transverse shift in the peak of the correlation images. To incorporate this we include phase parameters $\phi_i$ and $\phi_j$ in our model that are allowed to freely vary. From Eq.~\ref{eq:constant grid lines} and~\ref{Eq: analytical expression for Isg}, we compute the auto-correlation $I_{g} * I_{g}$ as
\begin{equation} \label{eq: analytical expression for grid-only auto correlation}
\begin{split}
    I_{g*g}(i,j) = &p^2 \Bigg(  \left( 1 - \frac{3}{4}\alpha\right)^2+\frac{1}{32} \alpha^2
    \cos \left(\frac{2 \pi }{p} i+ \phi_i\right) +\frac{1}{32} \alpha^2 \cos \left(\frac{2 \pi
     }{p}j + \phi_j\right) \\ & + \frac{1}{128} \alpha^2 \cos \left(\frac{2 \pi  }{p}(i-j)+\phi_i-\phi_j\right)+\frac{1}{128}
    \alpha^2 \cos \left(\frac{2 \pi  }{p}(i+j)+\phi_i+\phi_j\right) \Bigg),
\end{split}
\end{equation}
and the cross-correlation $I_{g} * I_{sg}$ as
\begin{equation} 
\begin{split}
    I_{g*sg}(i,j) = & p^2T\Bigg( \left(1 - \frac{3}{4}\alpha\right)^2+\frac{1}{32} \alpha^2
    e^{-a-b\cos\left( 2\theta\right)}\cos \left(\frac{2 \pi  }{p}i +\phi_i\right) \\& +\frac{1}{32} \alpha^2e^{-a+b\cos\left( 2\theta\right)} \cos \left(\frac{2 \pi
    }{p}j + \phi_j\right) + \frac{1}{128} \alpha^2 e^{-2a-2b\sin\left( 2\theta\right)}\cos \left(\frac{2 \pi  }{p}(i-j) +\phi_i - \phi_j\right)  \\ &+\frac{1}{128}
   \alpha^2 e^{-2a+2b\sin\left( 2\theta\right)}\cos \left(\frac{2 \pi  }{p}(i+j) +\phi_i+\phi_j\right) \Bigg),
   \label{eq:grid and sample-and-grid correlation2}
\end{split}
\end{equation}
where $(i,j)$ is the correlation space coordinate system, and $a$ and $b$ are defined above in Eq.~\ref{eq: a} and~\ref{eq: b}.
Both $I_{g*g}$ and $I_{g*sg}$ have matching trigonometric terms and the following general expression can be used to describe both
\begin{equation}
\begin{split}
    f(i,j) =  c_0 &+ c_1\cos \left(\frac{2 \pi  }{p}i + \phi_i\right)+c_2 \cos \left(\frac{2 \pi
    }{p}j + \phi_j\right)+c_3 \cos \left(\frac{2 \pi  }{p}(i-j)+ \phi_i- \phi_j\right) \\&+c_4 \cos \left(\frac{2 \pi  }{p}(i+j) + \phi_i+ \phi_j\right),
    \label{eq:correlation-fit-fuction}
\end{split}
\end{equation}
where $c_n$ represents the coefficient for each term. 
By computing $I_{g*g}$ and $I_{g*sg}$ we can then fit Eq.~\ref{eq:correlation-fit-fuction} to the resulting images in correlation space to determine the $c_n$ coefficients and extract the transmission and dark-field signals.

\section{Single-grid directional dark-field retrieval algorithm}

Experimental images captured on a pixelated detector can be represented as the discrete versions of the models $I_g$ and $I_{sg}$. Assuming that a local correlation of experimental images can be well-modelled by Eq.~\ref{eq: analytical expression for grid-only auto correlation} and~\ref{eq:grid and sample-and-grid correlation2}, we can then fit Eq.~\ref{eq:correlation-fit-fuction} to both of these local correlation space images and retrieve the parameters $T$, $\theta$, $\sigma_x$ and $\sigma_y$. The algorithm presented below was first tested on multiple sets of simulated data and then applied to experimental single-grid image data collected using a synchrotron source. Figures have been used in this section to aid in the explanation of the process using experimental data. The data collection process and details of this data is contained in Sec.~\ref{sec: results}.

\subsection{Computing the correlation-space coefficients}
\begin{figure}[htbp]
\centering\includegraphics[width =4.9in]{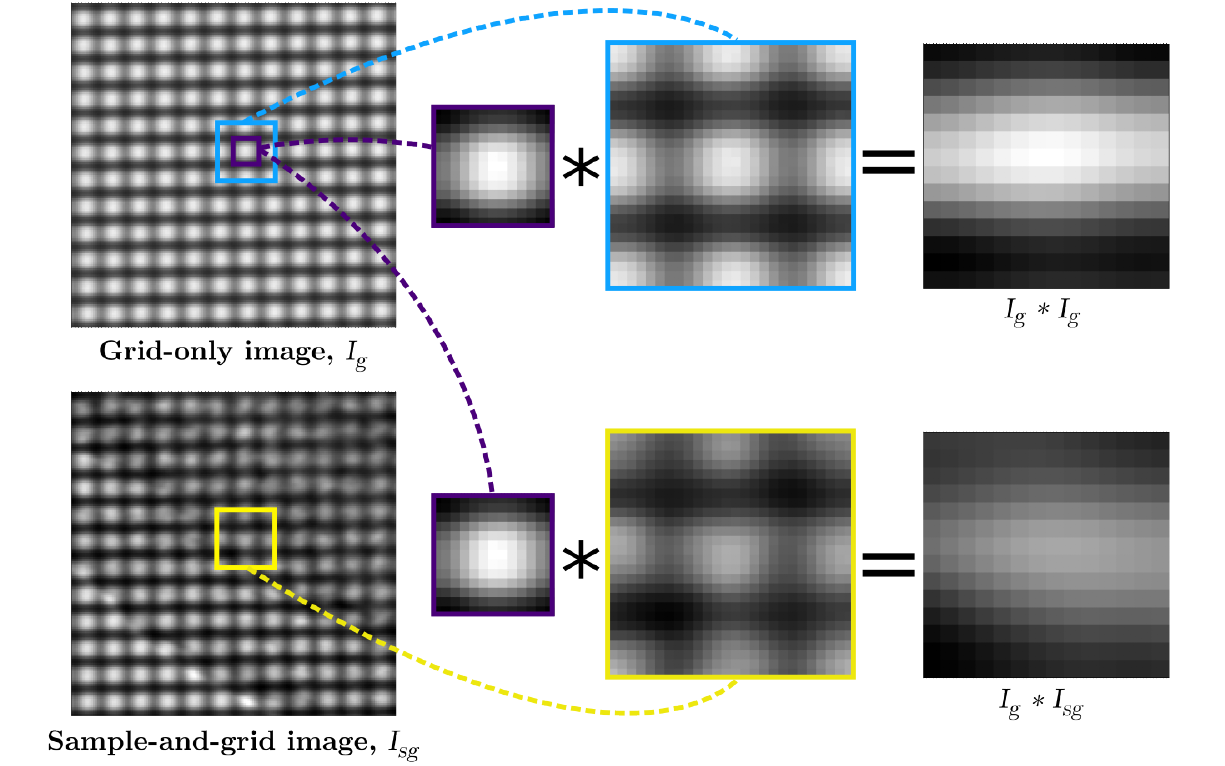}
\caption{A grid-only and a sample-and-grid image are shown on the left. Highlighted in purple is a selection for the $k \times k$ sized kernel. This kernel is correlated with the $2k \times 2k$ regions in the grid-only image (highlighted in blue) and sample-and-grid image (highlighted in yellow). The resulting auto-correlation image, $I_{g*g}$, and cross-correlation image, $I_{g*sg}$ are shown on the right. Note the reduced visibility and change in shape in the cross-correlation image compared to the auto-correlation image. These correlation-space images are fit with Eq.~\ref{eq:correlation-fit-fuction} to determine the dark-field scattering kernel parameters.  }
\label{fig: kernel and correlation}
\end{figure}

The correlation-space coefficients, $c_n$, depend on the dark-field scattering from each part of the sample and thus will vary across the image plane. In order to extract the local dark-field signal we select a kernel in the grid-only image of size $k \times k$, where $k$ is the kernel size in number of pixels. Typically $k$ is of order size $p$, the period of the reference pattern or grid. We then select another window of size $2k \times 2k$ centered around the kernel, also from the grid-only image. Taking the correlation of these two windows allows us to compute the local auto-correlation image $I_{g*g}$ (Eq.~\ref{eq: analytical expression for grid-only auto correlation}). Similarly we select a window at the same position of size $2k \times 2k$ in the sample-and-grid image to compute the local cross-correlation image $I_{g*sg}$ (Eq.~\ref{eq:grid and sample-and-grid correlation2}). We fit Eq.~\ref{eq:correlation-fit-fuction} to the auto-correlation image, providing values for the 5 coefficients that we denote with $c_{g*g,0}, c_{g*g,1}, ..., c_{g*g,4}$. Likewise with the cross-correlation image we get $c_{g*sg,0}, c_{g*sg,1}, ..., c_{g*sg,4}$. The selection of these windows is illustrated in Fig.~\ref{fig: kernel and correlation}. Computing these correlations locally is advantageous as it does not require the reference intensity pattern to be consistent in period and visibility across the whole image plane, giving it broad applicability to speckle reference patterns or grids with imperfections.

This correlation can be computed centred on each pixel within the image, but given the blurring is seen on a scale of the grid period, the final spatial resolution of the extracted dark-field image is reduced to the period of the grid. One option would be to only calculate a cross-correlation once per grid period. However, we found that as the grid period was not an integer number of pixels, this meant there was a variation in which part of the grid was selected in the kernel, leading to a periodic dependence of the correlation results on the kernel selected. To reduce the effect of this we moved the kernel and the corresponding comparison windows along one pixel at a time, computing a correlation result for every pixel position. These results were then averaged across grid-period-sized regions, to produce one image for each correlation coefficient (e.g. $c_{g*sg0}$). As we used an absorption grid with consistent period across the image plane, rather than simply using $k = p$, we tested which value for $k$ produced the most uniform values for each of $c_{g*g,n}$ across the image plane. As $I_g$ has no sample in it, then ideally the coefficients should be constant across the whole image plane, thus we use the value for $k$ that gives results closest to this. The combination of these two ideas provided correlation coefficient images with minimal variation induced by the kernel selection.

\subsection{Solving for the dark-field scattering kernel parameters}

The correlation coefficients are then compared to the analytical solutions given by Eqs.~\ref{eq: analytical expression for grid-only auto correlation} and~\ref{eq:grid and sample-and-grid correlation2} and solved as a set of simultaneous equations. Taking the ratios of each of these correlation coefficients pairs, e.g. $c_{g*sg,0}/c_{g*g,0}$, gives five equations that can be used to solve for four unknowns: $T$, $\theta$, $\sigma_x$ and $\sigma_y$. It is straightforward to first solve for the sample transmission
\begin{equation}
    T = \frac{c_{g*sg0}}{c_{g*g0}},
\end{equation}
which is shown in Fig.~\ref{fig:dark-field metrics}(c) for our test experimental dataset. The remaining ratio pairs can be rearranged such that the known values are on the left-hand-side of the expression, and the dark-field scattering parameters are on the right-hand-side. We then define each with a new variable $A_n$. For example, $A_1$ relates to the results from the $c_{g*sg,1}/c_{g*g,1}$ ratio and is defined as
\begin{equation}
   A_1 = \frac{p^2}{\pi^2} \ln\left(\frac{c_{g*sg1}}{T c_{g*g1}} \right)= - \sigma_x^2-  \sigma_y^2 - \left( \sigma_x^2 - \sigma_y^2\right) \cos \left( 2 \theta \right).
   \label{eq: A1}
\end{equation}
The value of $A_n$ can be computed as $p$, $T$ and $c_{g*sg,n}/c_{g*g,n}$ are known. These equations are combined into matrix form
\begin{equation}
    \begin{bmatrix}
        -1-\cos{2\theta} & \cos{2\theta} -1\\
        \cos{2\theta} -1 & -1-\cos{2\theta} \\
        -1 -\sin{2\theta} & \sin{2\theta} -1 \\
        \sin{2\theta} - 1 & -1 -\sin{2\theta}
    \end{bmatrix}
    \begin{bmatrix}
        \sigma_x^2 \\
        \sigma_y^2
    \end{bmatrix}
    =
    \begin{bmatrix}
        A_1 \\
        A_2 \\
        A_3 \\
        A_4 
    \end{bmatrix}.
    \label{eq:matrix simultanious equations}
\end{equation}
Using the Moore-Penrose inverse~\cite{moore_1920, penrose_1955} and the least squares method, Eq.~\ref{eq:matrix simultanious equations} is used to solve for a least-squares solution to $\sigma_x^2$ and $\sigma_y^2$ with the following result
\begin{equation}
    \sigma_x^2 = \frac{1}{8} \left( -A_1(2\cos2\theta +1) + A_2(2\cos2\theta -1) -A_3(2\sin2\theta + 1) +A_4(2\sin2\theta -1) \right),
\end{equation}
\begin{equation}
    \sigma_y^2 = \frac{1}{8} \left( A_1(2\cos2\theta -1) - A_2(2\cos2\theta +1) +A_3(2\sin2\theta - 1) -A_4(2\sin2\theta +1) \right).
\end{equation}
The set of equations in Eq.~\ref{eq:matrix simultanious equations} can also be rearranged to solve for $\theta$
\begin{equation}
    \theta = \frac{1}{2} \tan ^{-1} \left(  \frac{A_4-A_3}{A_2 - A_1} \right).
\end{equation}
In the case that $\sigma_x^2$ or $\sigma_y^2$ is a negative value, this indicates that the reference pattern has increased in visibility (a sharpening of the image) along that axis and this results in a complex scattering angle. Reasons that this may occur are discussed throughout Sec.~\ref{sec: results}. 

The solutions for these parameters that describe the dark-field scattering kernel can be expressed equivalently in two different ways. For any dark-field scattering kernel, the values of $\sigma_x$ and $\sigma_y$ can be swapped given an appropriate adjustment of $\theta$ to produce the same Gaussian kernel. Because of this we instead describe the standard deviation of the dark-field scattering kernel along the semi-major, $\sigma_M$, (maximum blur) and semi-minor, $\sigma_m$, (minimum blur) axes, and define $\theta$ to be the clockwise rotation of $\sigma_M$ from the positive $y$-axis (vertical direction) so that it describes the dominant scattering direction by representing the rotation of the semi-major axis. To do this, we apply the following conditional transforms to the numerical solutions 
\begin{equation}
   \text{if } \sigma_x^2 < \sigma_y^2 \rightarrow \sigma_M = \sigma_y, \sigma_m = \sigma_x,
\end{equation}
\begin{equation}
    \text{if }\sigma_y^2 < \sigma_x^2 \rightarrow \sigma_M = \sigma_x, \sigma_m = \sigma_y, \theta \mathrel{+}= \pi/2,
\end{equation}
\begin{equation}
    \text{if } \theta < 0 \rightarrow \theta \mathrel{+}= \pi.
\end{equation}
\begin{equation}
    \text{if } \theta > \pi \rightarrow \theta \mathrel{-}= \pi.
\end{equation}
These transforms result in unique solutions for the dark-field scattering kernel that are therefore consistent across the image and $\theta$ is mapped between $0$ and $\pi$. 

\subsection{Quantitative dark-field metrics}
\begin{figure}[htbp]
\centering\includegraphics[width=4.9in]{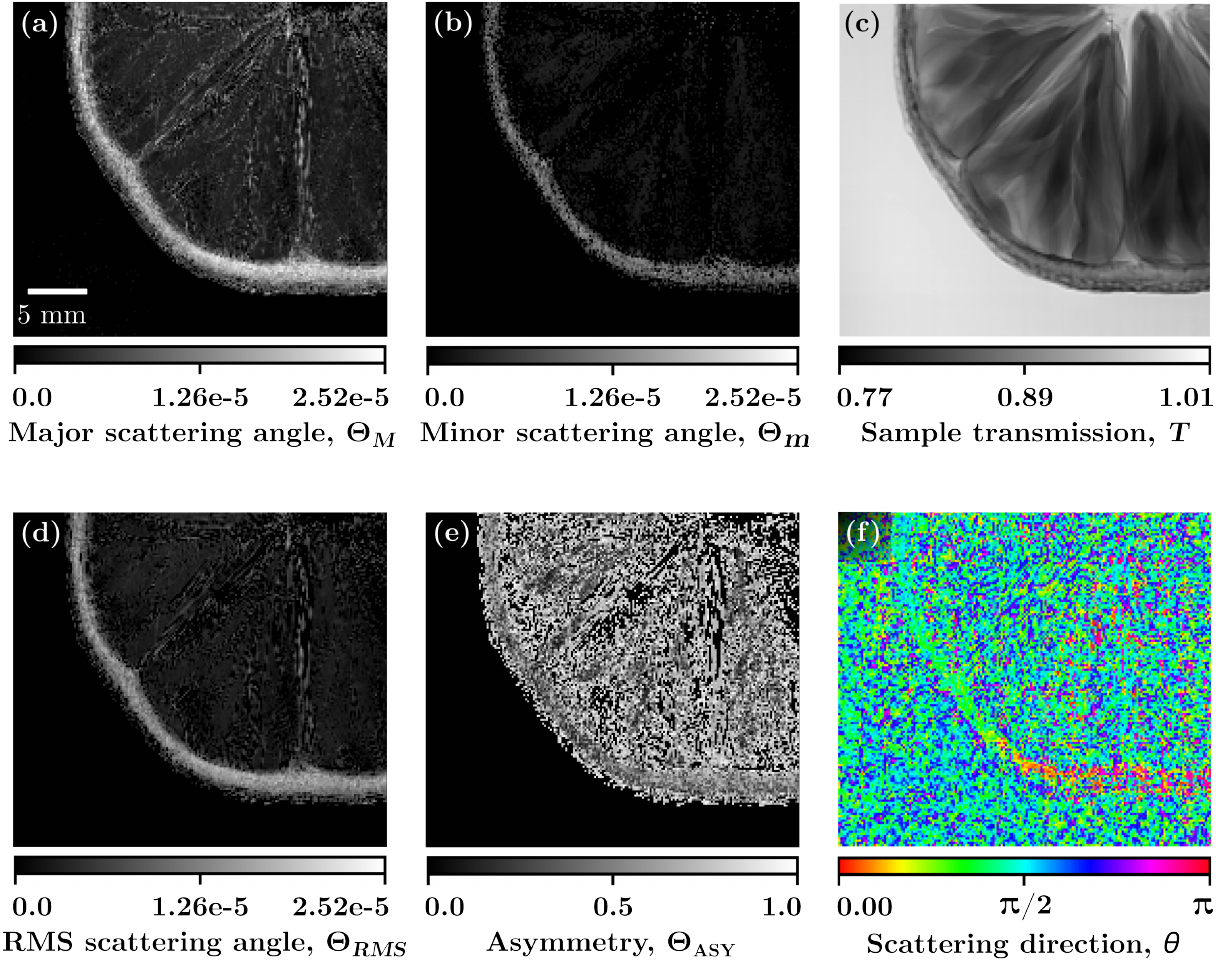}
\caption{Various dark-field signal measures of a lemon slice retrieved using the single-grid directional dark-field retrieval algorithm. (a) The major scattering angle, $\Theta_M$, in radians. (b) The minor scattering angle, $\Theta_m$, in radians. (c) The retrieved sample transmission T. (d) The dark-field strength characterised as the root-mean-square of the scattering angles, in radians. (e) The dark-field asymmetry characterised by a measure of the difference between the scattering angles. (f) The dominant scattering direction in radians, where $\theta = 0 \text{ or } \pi$ corresponds to vertical scattering (red) and $\theta = \pi/2$ corresponds to horizontal scattering (blue).}
\label{fig:dark-field metrics}
\end{figure}

The dark-field parameters extracted using the above retrieval algorithm can be used to create meaningful metrics that tell us about the dark field of the sample. The solutions for the dark-field scattering widths, $\sigma_M$ and $\sigma_m$, are dependent on the experimental setup, primarily the object-to-detector distance, $ODD$. Here we transform these parameters into quantitative measures of the dark-field signal. The scattering widths are transformed into semi-minor and semi-major scattering cone half-angles using the small angle approximation
\begin{equation}
    \Theta_m = \frac{\sigma_m}{ODD},
\end{equation}
\begin{equation}
    \Theta_M =  \frac{\sigma_M}{ODD},
\end{equation}
where $\Theta_M$ describes the larger cone half-angle that the sample has scattered the beam and $\Theta_m$ is the cone half-angle perpendicular to $\Theta_M$ and is also the smaller scattering angle from that part of the sample. Contained within a cone described by $\Theta_M$ and $\Theta_m$ is $\sim 68.2\%$ of the initial beamlet intensity. Experimental results for the scattering cone half-angles produced by this single-grid directional dark-field retrieval algorithm are shown in Fig.~\ref{fig:dark-field metrics}. Complex values for these scattering angles correspond to a sharpening of the image, or a "negative blur", so when presenting them in images throughout this paper we display the values as zero, however we treat the values as complex in all calculations.

Previous measures of the dark-field signal quantify the strength, or degree of scattering, by a change in visibility of the grid lines. As we can directly measure the scattering angle using this algorithm we are able to define a quantitative measure of the dark-field strength as the root-mean-square of the scattering angles
\begin{equation}
    \Theta_{RMS} = \sqrt{\frac{\Theta_M^2 + \Theta_m^2}{2}}.
\end{equation}
When $\Theta_M$ and $\Theta_m$ are small $\Theta_{RMS} \rightarrow 0$ signifying minimal scattering. As $\Theta_M$ and $\Theta_m$ become larger so does $\Theta_{RMS}$. This scattering angle describes the approximate circular cone that contains $68.2\%$ of the beamlet if the scattering were symmetrical about the initial direction of the beamlet. It is possible for $\Theta_{RMS}$ to be a complex value, indicating insufficient sample-induced blur to overcome the sharpening introduced in the experiment.

The asymmetry, or how directional the dark-field signal is, can be measured by the relative difference between the scattering angles. However, weakly scattering samples, or samples that have a strong dominant scattering direction may have complex scattering angle(s). As the asymmetry is integral to meaningfully interpret other dark-field parameters (as demonstrated in Sec.~\ref{sec: HSV}) it is important for this metric to be able to function in the presence of complex values. To do this we only measure the asymmetry for sufficiently strong signals where $\Theta_{RMS} \in \mathbb{R}$. Thus, requiring that $\Theta_M \in \mathbb{R}$ and $|\Theta_M|\geq|\Theta_m|$. Below is a continuous piecewise function that quantifies the asymmetry of the sample scattering given $\Theta_M \in \mathbb{R}$ but allowing for $\Theta_m$ to be complex.
\begin{equation}
    \Theta_{ASY} =  \left\{
    \begin{array}{ll}
          1 - \frac{\Theta_m}{\Theta_M} & \frac{1}{9} \leq \left(\frac{\Theta_m}{\Theta_M}\right)^2 \leq 1, \\
          \\
          \sqrt{\frac{\Theta_{M}^2-\Theta_m^2}{2\Theta_M^2}} & -1 \leq \left(\frac{\Theta_m}{\Theta_M}\right)^2 < \frac{1}{9},\\
          \\
          0 & \text{Otherwise.}
    \end{array} 
    \right.
\end{equation}
Here $\Theta_{ASY} = 0$ means the dark-field scattering is rationally symmetric in the $xy$ plane as $\Theta_M = \Theta_m$. The case when $\Theta_{ASY} \rightarrow 1$ means the dark field is becoming very asymmetrical and predominately only scattering along one direction in the image plane. When $(\Theta_m/\Theta_M)^2 < -1$ this indicates there is very low dark field induced blur with $\Theta_{RMS} \notin \mathbb{R}$. Thus, it is not relevant to consider the asymmetry of the dark-field scattering kernel at that location in the image plane. The direction of asymmetrical scatter is described by the dominant scattering direction, $\theta$, that is calculated as one of the parameters defining the dark-field scattering kernel. These dark-field metrics are all shown in Fig.~\ref{fig:dark-field metrics}.

\subsection{Hue-saturation-value images of the dark field}\label{sec: HSV}

\begin{figure}[htbp]
\centering\includegraphics[width=4.9in]{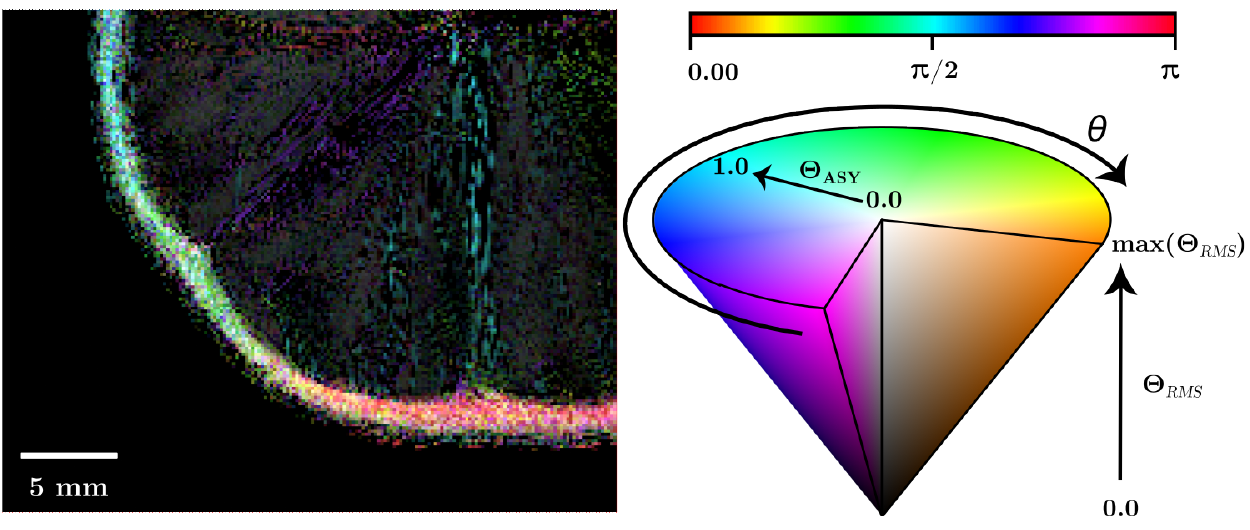}
\caption{Dark-field HSV image of a lemon slice. Hue (colour) represents the dominant dark-field scattering direction, $\theta$ within the $xy$ image plane. A value of $\theta = 0 \text{ or } \pi$ radians corresponds to predominantly vertical scattering (red) and $\theta = \pi/2$ radians corresponds to predominantly horizontal scattering (blue). Saturation represents the dark-field asymmetry, where the pixels will appear in full colour at high asymmetries or with low saturation (gray scale) when the scattering is isotropic and the directional information has less relevance. Value (brightness) represents the dark-field strength with brighter pixels denoting stronger scattering. For this image the brightest pixels have $\Theta_{RMS} = 16~\upmu$rad. It can be seen that the lemon's pith (white interior of the rind) has the strongest scattering and scatters the x-ray beam perpendicular to the surface of the lemon indicating the pith contains elongated microstructures that are aligned parallel to the surface. Edges of the juice sacs and segment walls produce a weaker dark-field signal.}
\label{fig: Lemon HSV}
\end{figure}

The dark-field metrics can be combined to create a meaningful image that represents the dark-field signal in an easy-to-interpret manner. Hue-Saturation-Value (HSV) images can be used to combine three different `gray-scale' images into one colour image. This is demonstrated in Fig.~\ref{fig: Lemon HSV} which combines the three dark-field metrics. The dominant scattering direction, $\theta$, is used to determine the hue (colour) of each pixel. The asymmetry, $\Theta_{ASY}$, is used for saturation so that isotropically-scattering areas of the image appear with less colour, as the scattering direction information is less relevant. The dark-field strength,
$\Theta_{RMS}$, indicates the value or brightness of the pixel, so regions that are highly scattering appear bright in the image and weakly-scattering regions are dark. 

\section{Retrieval algorithm experimental tests} \label{sec: results}

The single-grid directional dark-field retrieval algorithm detailed above allows us to extract sub-pixel sample structure information from a single exposure of the sample. Below we test the robustness of this retrieval algorithm with data collected in hutch 3B of the Imaging and Medical Beamline at the Australian Synchrotron. Hutch 3B has a typical source-to-sample distance of 135 m. A commercially available absorption grid with 90 $\upmu$m holes and 65 $\upmu$m thick grid lines (Essa ISO 3310 ASTM stainless steel geological sieve) was placed 62 cm upstream of the fixed sample position. Various detector positions were used, with an object-to-detector distance of either 50 cm, 100 cm or 150 cm, achieved by moving the detector further downstream of the sample, and giving a grid period of $12.48$, $12.55$ and$ 12.58$ pixels respectively. The images were collected using the "Ruby" detector, a lens-coupled scintillator (25 $\upmu$m Gadox) with an sCMOS sensor producing an effective pixel size of 12.3 $\upmu$m. See Fig.~\ref{fig: experimental set up} for schematic of experimental setup. Data presented in this paper has an exposure time of 30 s unless specified otherwise, and all imaging was done with an x-ray energy of 25 keV. Time-resolved dark-field imaging was all conducted and experimental details for this are presented in Sec~\ref{sec:time}.

The samples used included a lemon slice (Fig.~\ref{fig: kernel and correlation},~\ref{fig:dark-field metrics},~\ref{fig: Lemon HSV},~\ref{fig: phase figure}), thin pieces of Medium Density Fibre (MDF), Tasmanian oak and balsa wood (Fig.~\ref{fig: ODD compariosn}), carbon fibres threaded onto a plastic sheet (Fig.~\ref{fig: experimental set up},~\ref{fig: noise compariosn}, Visualisation 1) and carbon fibres wound around a plastic tube (Fig.~\ref{fig: phase figure}). For each sample we collected a sample-only image, $I_s$, and then the grid was moved into the beam to collect the sample-and-grid image, $I_{sg}$. Finally the sample was removed from the beam and the grid-only image, $I_g$, was collected. The order of image collection $I_s \rightarrow I_{sg} \rightarrow I_g$ is important to ensure the sample and grid remain in the exact same position between each image. While the sample-only image is not required for the retrieval algorithm it was collected to demonstrate how the dark-field and phase signals are complementary as discussed in Sec.~\ref{sec: phase results}. Flat field and detector dark-current images were collected to allow for flat and dark correction of the experimental images. 

The dark-field parameters $\theta$, $\Theta_M$ and $\Theta_m$ in our dark-field model should be independent of the experiment setup. However, as the experimental images are discretised versions of these models our ability to retrieve the dark field is limited by how well we can resolve the changes in the reference intensity pattern. Changes in the reference intensity can also be affected by other sources of blur or sharpening from the experimental setup. In order to determine the robustness of the retrieval algorithm we collected data with different object-to-detector distances, $ODD$, (Sec.~\ref{sec: ODD}) and with different exposure times (Sec.~\ref{sec: noise}). In Sec.~\ref{sec: phase results} we present results that demonstrate the complementarity of the directional dark field to the differential phase images that can be extracted from the same data set. Finally we present time-resolved directional dark-field imaging in Sec.~\ref{sec:time}.

\subsection{Varying object-to-detector distance} \label{sec: ODD}

\begin{figure}[htbp]
\centering\includegraphics[width=4.9in]{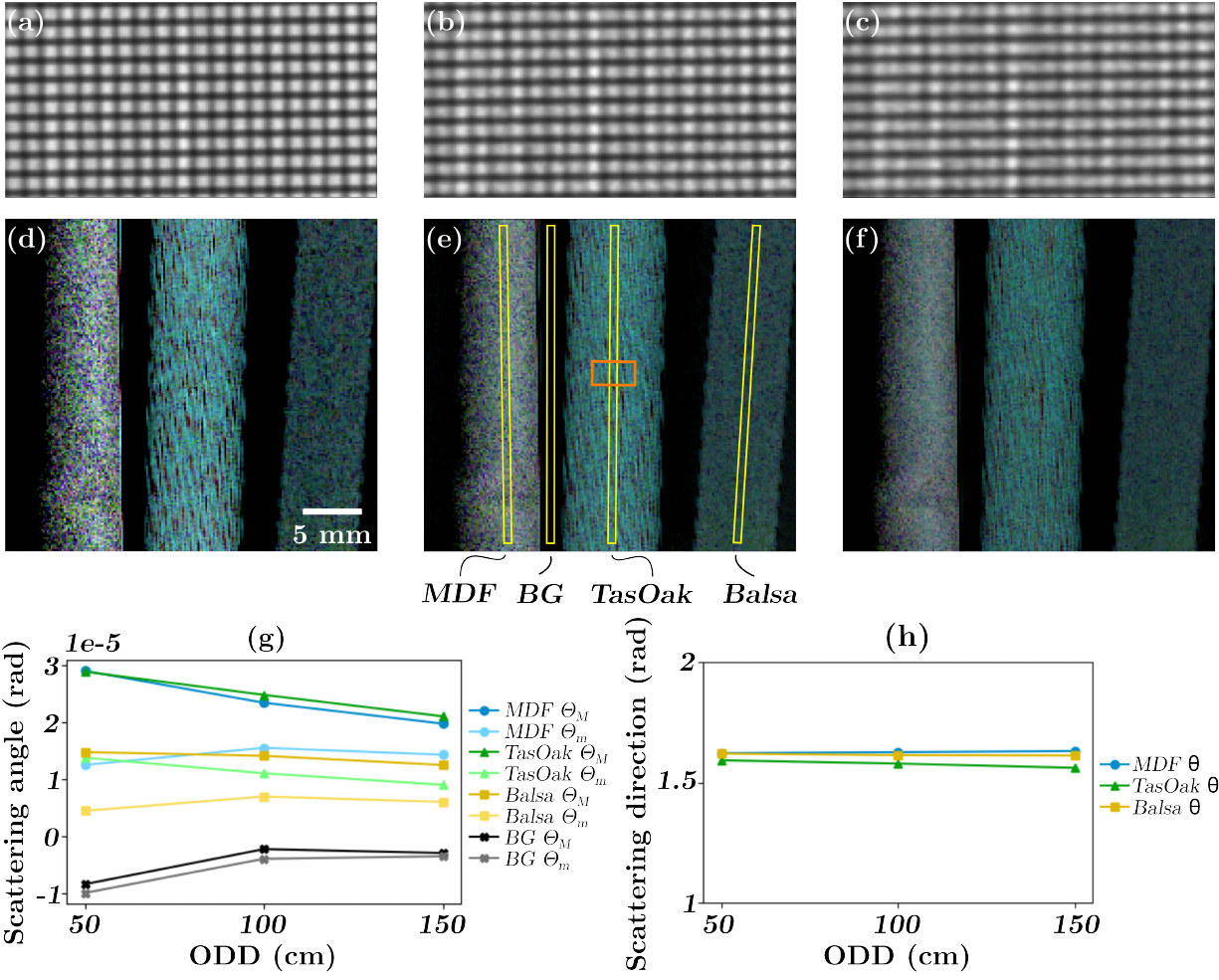}
\caption{Data collected from samples of MDF, Tasmanian oak and balsa. Zoomed in sections (shown in orange in (e)) of the sample-and-grid image, and dark-field HSV images, are shown at object-to-detector distances of 50~cm (a,d), 100~cm (b,e) and 150~cm (c,f). See Fig.~\ref{fig: Lemon HSV} for colour wheel, and note HSV images in this figure are scaled using the same max$(\Theta_{RMS})$ value of $40~\upmu$rad. Regions of interest from each sample and a background section of the image were selected for analysis as shown in (e). (g)~The average scattering angles, $\Theta_M$ and $\Theta_m$, for each of the regions of interest. Note that complex scattering angles are displayed on this plot as negative real values. (h) The dominant scattering direction, $\theta$, vs the object-to-detector distance. Based on the near vertical alignment of the wooden fibres it is estimated that $\theta$ for the Tasmanian oak and balsa wood samples would be $1.60$ and $1.63$ rad respectively. The texture seen in the Tasmanian Oak sample is produced by resolved phase fringes affecting the dark-field reconstruction, see Sec. \ref{sec: phase results}. }
\label{fig: ODD compariosn}
\end{figure}

We imaged three different wooden samples; MDF, Tasmanian oak, and balsa wood, at 50 cm, 100 cm and 150 cm $ODD$. Fig.~\ref{fig: ODD compariosn} shows examples of the sample-and-grid images, the dark-field HSV images, as well as plots of the scattering angles, $\Theta_M$ and $\Theta_m$, and the dominant scattering direction, $\theta$, each as a function of $ODD$. Fig.~\ref{fig: ODD compariosn}(g) shows how the scattering angles vary with $ODD$, understanding why there is a variation is important for ensuring this technique correctly extracts the scattering angles. The scattering angles for the background region are all complex (shown as negative real values in Fig. \ref{fig: ODD compariosn}(g)), this indicates a sharpening of the background region. Upon comparison of the grid-only and sample-and-grid images it was noted that there was a slight absolute decrease in intensity in background regions of the sample-and-grid images. We hypothesise that as the sample has been introduced, it has attenuated some of the x-ray beam and as a result the point-spread-function of the scintillator spreads less overall intensity across the whole image in comparison to the grid-only image. This reduction in intensity reduces the mean of the sample-and-grid image, but does not alter the amplitudes of the reference pattern. This increases the visibility of the reference pattern resulting in an effective-sharpening.
As this effective-sharpening occurs at the scintillator (a fixed distance from the detector), then the complex blurring widths remain constant regardless of the detector position and thus the associated complex scattering angles become closer to $0i$ radians at larger $ODD$. 

The scattering angles of the sample regions tend to decrease with increased object-to-detector distance. Other sources of blur such as Compton scatter need to be considered as they will be interpreted as dark-field scattering by this algorithm. The scattering angles associated with Compton scatter in general are much larger than that of the dark field and so at a short distance could appear as local blurring but at a longer distance fewer of the Compton scattered photons would reach the detector and so this effect may appear more as attenuation of the beam rather than blur. If so, at greater $ODD$ there would be less blur due to wide-angle scattering mechanisms such as Compton scattering. We also see a much weaker dark-field signal at short ODD, which makes the dark-field retrieval less robust.

We see that the Tasmanian oak and balsa wood both have a large difference in the major and minor scattering angles at all distances, indicating that they contain elongated wooden fibres that anisotropically scatter the x-ray beam. However, as MDF is made of many small fibres glued together, the x-ray wavefield scatters in a variety of directions. At short object-to-detector distances small local regions have different dominant scattering directions, as can be seen as the variety of hues in~\ref{fig: ODD compariosn}(d). As the detector is moved further back the blurring from these local regions begins to overlap at the image plane creating a more homogeneous scatter that brings the semi-major and semi-minor scattering angles closer together.

The dark-field scattering kernel's blurring widths, $\sigma_M$ and $\sigma_m$, are directly proportional to $ODD$. The non dark-field factors that introduce blur/sharpening seem to have a larger affect at shorter distances. These preliminary results indicate that careful selection of a large object-to-detector distance, such that the grid remains visible (not completely blurred out) and dark-field signal is sufficiently greater then other blur/sharpening signals is required for quantitative scattering angle retrieval. Further investigation of this would be beneficial.

Regardless of object-to-detector distance, the dominant scattering direction, $\theta$, is very consistent, as shown in Fig.~\ref{fig: ODD compariosn}(h). Thus the single-grid directional dark-field retrieval algorithm provides a robust solution to determining the dominant scattering direction, independent of experimental setup.

\subsection{Varying exposure time} \label{sec: noise}

\begin{figure}[htbp]
\centering\includegraphics[width=4.9in]{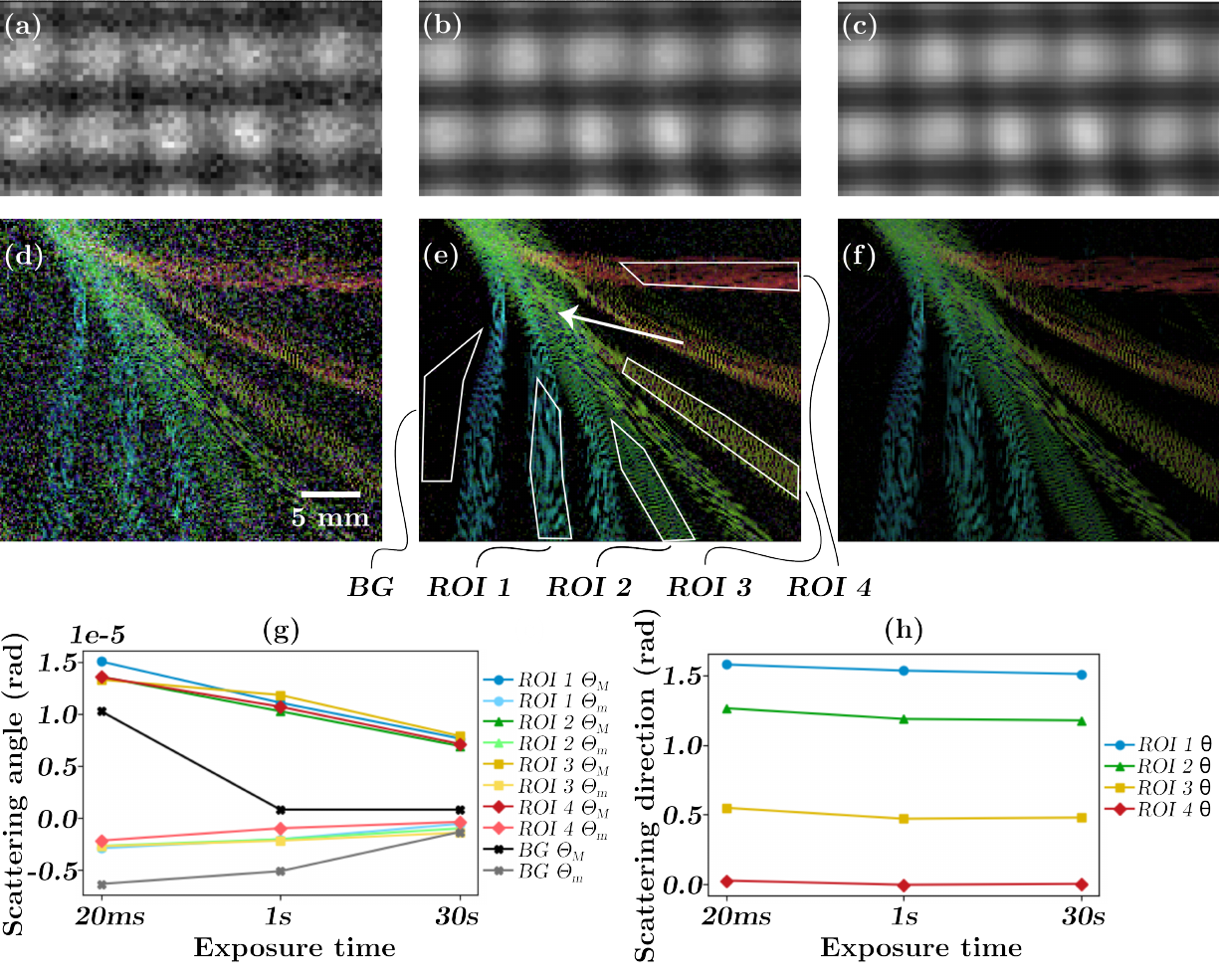}
\caption{Data collect of carbon fibre bundles oriented in different directions. Zoomed in section (location shown by white arrow in (e)) of the sample-and-grid image, and dark-field HSV images, with effective exposure times of 20 ms (a,d), 1 s (b,e) and 30 s (c,f). See Fig.~\ref{fig: Lemon HSV} for colour wheel, and note all images in this panel are scaled using the same max$(\Theta_{RMS})$ value of $24~\upmu$rad. Four sample regions of interest and a background section of the image were selected for analysis as shown in (e). (g) The average scattering angles, $\Theta_M$ and $\Theta_m$, for each of the regions. Note that complex scattering angles are displayed on this plot as negative real values. (h) The average dominant scattering direction, $\theta$, within a region vs the effective exposure time. Based on the orientation of the fibres in the image plane we estimate the expected dominant scattering directions to be $\theta_{ROI1} = 1.63$, $\theta_{ROI2} = 1.07$, $\theta_{ROI3} = 0.56$ and $\theta_{ROI4} = 0.01$ rad. The texture seen in the carbon fibre bundles is produced by resolved phase fringes affecting the dark-field reconstruction, see Sec. \ref{sec: phase results}.}
\label{fig: noise compariosn}
\end{figure}

We captured multiple images of a sample at various exposure times. A lower exposure time means more noise is contained in the image. This makes it more difficult to determine subtle changes in the reference intensity pattern. Bundles of carbon fibre threads were extracted from carbon fibre cloth and sewn onto a thin plastic sheet to hold them in position with different orientations. The bundles contain threads that run parallel to each other in one direction. We collected 20 ms and 1 s exposures of the sample-and-grid image, and analysed retrieved images with effective exposure times of 20 ms, 1 s, and 30 s (the average of 30 images with 1 s exposure times). The flat field, dark current and grid-only images were collected with a longer exposure time, so that we had relatively noise free images for flat and dark correction and a smooth reference pattern image for comparison. This is appropriate as it does not increase the radiation exposure to the sample or affect the frame rate used to collect the sample-and-grid images. Fig.~\ref{fig: noise compariosn} shows example sample-and-grid image, the HSV images for each exposure time, and plots of the scattering angles and the dominant scattering direction. Fig.~\ref{fig: noise compariosn}(g) shows that at lower exposure times the major and minor scattering angles diverge from each other. We hypothises this is due to the increased noise in the image resulting in an effective blur or sharpening of the grid and thus altering the average value for $\Theta_M$ and $\Theta_m$ respectively. This can also be seen in the background regions of the image with lower exposure time. As the exposure time is increased the background scattering angles begin to converge to $0$, as do the minor scattering angle for each of the region of interest (ROI). It is expected that carbon fibres will scatter very strongly in one direction and so the scatter in the minor direction will be minimal. As the exposure time is increased the major scattering angle decrease as the retrieval becomes less affected by image noise. 

Importantly, the major scattering angle of the sample remains above the noise threshold and this allows for the dominant scattering direction to be determined from low exposure images as shown in Fig.~\ref{fig: noise compariosn}(h). This result is important as it implies that directional information about a sample can be retrieved in the presence of high image noise. This will allow for low dose dynamic imaging of samples where directional information is the measure of interest. 

\subsection{Dark field and phase contrast comparison} \label{sec: phase results}
The results from the single-grid directional dark-field retrieval algorithm are complementary to that of a single-grid phase retrieval method. The dark-field signal is the blur introduced in an image due to multiple sub-pixel structures altering the phase of the x-ray wavefield or inducing small angle x-ray scattering (SAXS/USAXS). This contrasts with resolved phase effects that result in reference pattern shifts. Local intensity variations from sample absorption or propagation-based bright-dark fringes around the edges of resolved structures can alter the intensity of the reference pattern in a way that appears like a blurring, sharpening or shift of intensity. This is particularly apparent when the scale of the attenuation or propagation-based phase contrast is similar in size to the local period of the reference pattern, resulting in artefacts in the retrieved dark-field signal. The wooden samples shown in Fig.~\ref{fig: ODD compariosn} and the carbon fibre bundles in Fig.~\ref{fig: noise compariosn} show this type of artefact as black regions or stripes in the HSV images. This occurs when the intensity is changed such that it looks like an increase in visibility. 

\begin{figure}[htbp]
\centering\includegraphics[width=4.9in]{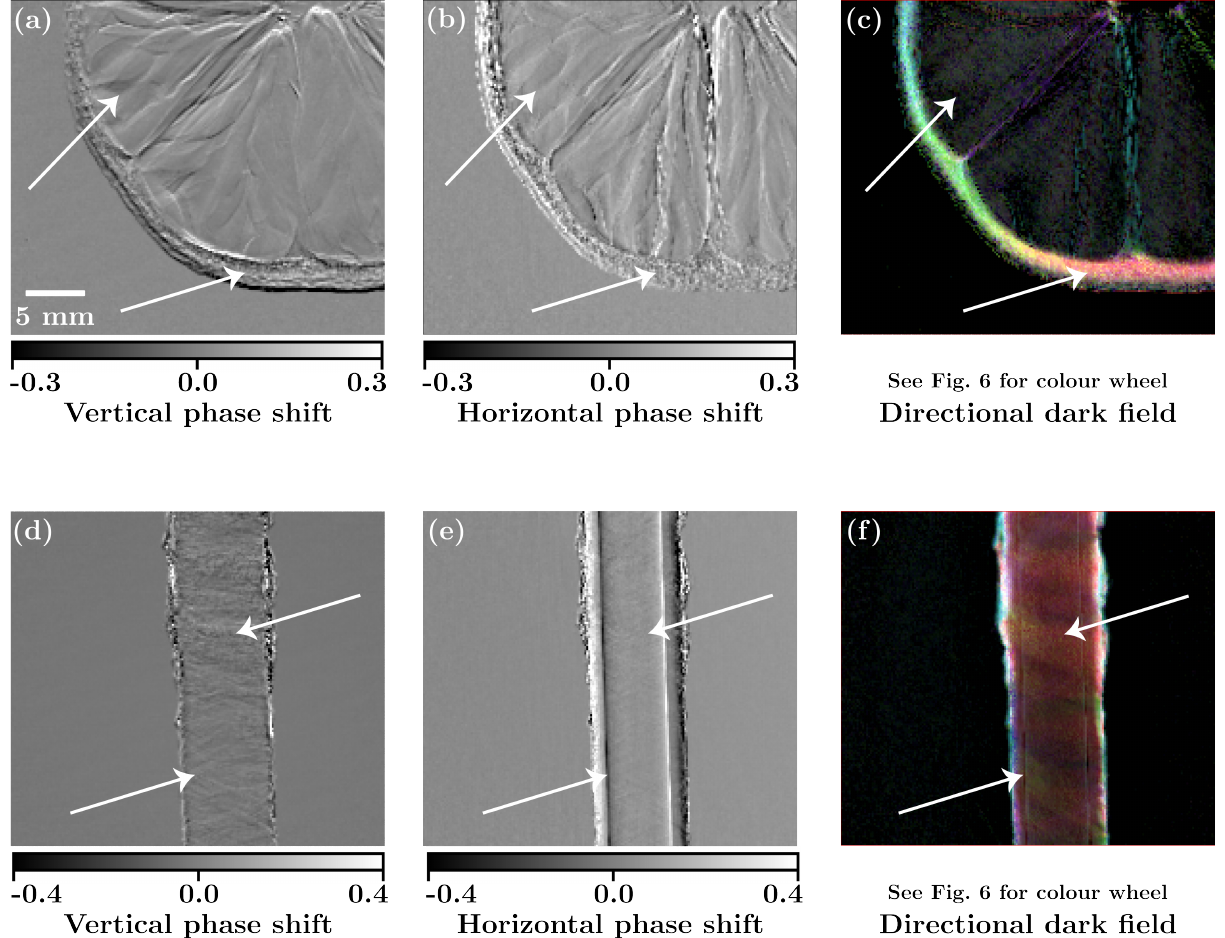}
\caption{Vertical phase shift, horizontal phase shift and directional dark-field HSV images for a lemon slice (a), (b) and (c), and a plastic tube wrapped in carbon fibre thread, (d), (e) and (f) respectively. Imaged at 150 cm $ODD$. Phase shifts are given in units of how far the grid pattern has been transversely shifted across the detector, in pixels. See Fig.~\ref{fig: Lemon HSV} for HSV colour wheel, (c) and (f) have a max$(\Theta_{RMS})$ value of $16~\upmu$rad and $25~\upmu$rad respectively. Note the strong dark-field signal in the lemon pith (white fibrous structure between the rind and juice cells) reveals that the pith has more microstructure than the rind, which is not obvious in the differential phase images and note the absence of a dark-field signal from the juice cells that are easily observed in the differential phase images. The strong differential phase signal for the internal edge of the plastic tube is also absent in the dark-field retrieval. Observe that the orientation of the wrapped carbon fibres can be seen in the dark-field image. }
\label{fig: phase figure}
\end{figure}

To help demonstrate that dark-field imaging complements phase imaging by providing new information, we can refine our data to reduce the attenuation and propagation-based phase artefacts by dividing the sample-and-grid image by the sample-only image. This leaves an image that contains only the grid with sample induced shift and blur-based distortions which are all accounted for in our dark-field model, and removes the propagation-based phase contrast intensity variations. The differential phase is extracted by using a cross-correlation method to determine the local shift of the grid lines in the image plane~\cite{morgan2011quantitative}. Fig.~\ref{fig: phase figure} shows the results of both the differential phase and directional dark field for a partially dried lemon slice and plastic tube wrapped in carbon fibre threads. Note how some features are more apparent in the different contrast modalities. Fig.~\ref{fig: phase figure}(c) can be compared to Fig.~\ref{fig: Lemon HSV} to note that without dividing out the sample-only image the dark-field signal is affected by some propagation-based phase contrast fringes, however the important features are all present. If a sample was static and would not suffer from twice the exposure time, then collecting a sample-only image would allow for a cleaner retrieval of both the dark field and the differential phase images of the sample. To apply this extra step during dynamic imaging, one possibility would be to place the sample upstream of the reference pattern, and introduce an optical element to split the beam prior to passing through the reference pattern and collect the relevant images on two detectors. An alternative would be an iterative process that takes the retrieved sample thickness (e.g. from Fig.~\ref{fig:dark-field metrics} (c)), numerically propagates this, and then divides out the simulated `sample-only' image at the given distance, similar to Groenendijk et al.~\cite{groenendijk2020}, but in this case dividing out the attenuation and phase effects from the sample-and-grid image.

\subsection{Time-resolved imaging of a changing sample} \label{sec:time}
To demonstrate that this method is capable of doing time-resolved imaging we collected data of an irreversible process. The carbon fibre sample shown in Fig.~\ref{fig: noise compariosn} was lowered into a small container of water such that only the ends of the fibres came in contact with the water. This resulted in capillary action pulling the water up into the bundles of carbon fibres. The retrieved sample transmission and directional dark-field images have been combined into a movie, see Visualisation 1 (\url{https://opticapublishing.figshare.com/articles/media/Time-resolved_directional_dark-field_imaging_of_carbon_fibres/22057232}).  As the water flows into the carbon fibres, as seen in the attenuation image, it fills up tiny pockets of air, and so the x-beam now travels through carbon fibre-water interfaces instead of carbon fibre-air interfaces. This decreases the angle of refraction occurring from sub-pixel microstructures and thus reduces the strength of the dark-field signal. This can be observed in Visualisation 1 as a reduction in brightness or the dark-field image with time. Similar to Fig. \ref{fig: noise compariosn} resolved phase effects have introduced artefacts resulting in dark regions within the carbon fibre bundles. However, the orientation of each bundle, and the reduction in dark-field strength with time, can still be clearly observed. 

This was imaged at the Micro-Computed Tomography Beamline at the Australian Synchrotron~\cite{mct2023}. Experimental Hutch B is located 25 m from a bending-magnet source. A commercially available absorption grid with 25 $\upmu$m holes and 30 $\upmu$m thick grid lines (Essa ISO 3310 ASTM stainless steel sieve) was placed 30 cm upstream of the sample position grid period of $8.5$ pixels. An object-to-detector distance of 140 cm was used. The images were collected using the "white-beam" detector with an effective pixel size of 6.5 $\upmu$m~\cite{mct2023}. Data was collected at a 25 Hz frame rate and 20 ms exposure time at an x-ray energy of 25 keV.

\section{Conclusion}
The single-grid directional dark-field retrieval algorithm presented above is a robust analysis technique that is able to extract directional scattering information about sub-pixel microstructures in a sample. Using a synchrotron source we demonstrated that it can be applied to a range of samples and can quantify the dominant scattering direction and the dark-field scattering angles in the semi-minor and semi-major directions of the scattering cone. These dark-field parameters can be converted into quantitative measures of the dark-field strength and asymmetry (Fig.~\ref{fig:dark-field metrics}), allowing for the production of easy-to-interpret hue-saturation-value images (Fig.~\ref{fig: Lemon HSV}). 

In order to determine the robustness of the algorithm we imaged samples at various object-to-detector distances and exposure times. We demonstrated that other sources of blur or sharpening can affect the measurement of the scattering angles, $\Theta_M$ and $\Theta_m$, and thus care needs to be taken to choose a sufficiently-large object-to-detector distance such that the dark-field signal is the dominant source of blurring (Fig.~\ref{fig: ODD compariosn}). As expected, low exposure times introduced noise into the image and affected the measurement of the scattering angles (Fig.~\ref{fig: noise compariosn}). Importantly, regardless of the $ODD$ or exposure time the algorithm was able to consistently extract the dominant scattering direction, $\theta$. Phase and attenuation-based intensity variations of similar local appearance to the reference pattern can produce artefacts in the retrieved dark-field images using this technique, although these can be minimised if a sample-only image is available. Alternatively future work could look into modelling a sample-only image to assist in the reduction of these artefacts.
The dark-field signal provides complementary information to the differential phase that can also be extracted from the same image set (Fig.~\ref{fig: phase figure}).

Dark-field imaging of dynamic samples is primarily restricted to slow~\cite{prade_observing_2015} or repeated dynamics~\cite{gradlDynamicVivoChest2019}. As the approach described here only requires one sample exposure, and can be applied to noisy images, this allows imaging of irreversible processes at high frame rates as demonstrated in Sec~\ref{sec:time}. This new single-grid directional dark-field retrieval algorithm expands on existing methods by providing a set-up-independent measure of the dark-field strength, and a measure of the dark-field asymmetry and direction. As a result it will now be possible to extract quantitative directional dark-field information from dose-sensitive or dynamic samples.

\section*{Acknowledgments}

The authors are grateful for help provided by the beamline scientists, Chris Hall, Daniel Hausermann, Anton Maksimenko and Matthew Cameron, at the Imaging and Medical beamline at the Australian Synchrotron, part of ANSTO, where the experiments reported in the main body of the paper were completed under proposal 18642. We also thank the beamline scientists, Andrew Stevenson and Benedicta Arhahtari, at the Micro-Computed Tomography beamline at the Australian Synchrotron, part of ANSTO, where the time-resolved experiment (Visualisation 1) was completed under proposal 19138. KM acknowledges support from the Australian Research Council (FT18010037) and MC acknowledges funding from the Australian Research Training Program (RTP). 

\section*{Data Availability Statement}
An example data set, code and in depth instructions for completing this analysis can be found at \url{https://github.com/Xray-grill/DiDaFi_MKCroughan}. The data underlying the results presented in this paper are not publicly available at this time but may be obtained from the authors upon reasonable request.

\bibliographystyle{unsrt}  
\bibliography{Manuscript}

\end{document}